\PassOptionsToPackage{numbers,sort&compress}{natbib}
\documentclass{article}
\usepackage[preprint]{tackling_climate_workshop_style}
\usepackage[T1]{fontenc}
\usepackage{hyperref}
\usepackage{url}
\usepackage{booktabs}
\usepackage{amsmath,amssymb}
\usepackage{graphicx}
\usepackage{microtype}
\hypersetup{hidelinks}

\title{Steering Diffusion Priors with Sparse Observations for High-Resolution Temperature Downscaling}
\author{%
\normalfont\small
\begin{tabular}{@{}c@{}c@{}c@{}}
\parbox[t]{0.32\textwidth}{\centering
  \textbf{Anirudh Avireddy}\\
  Shiv Nadar University, India} &
\parbox[t]{0.32\textwidth}{\centering
  \textbf{Manmeet Singh}\\
  Western Kentucky University, USA} &
\parbox[t]{0.32\textwidth}{\centering
  \textbf{Shivanshi Singh}\\
  Ashoka University, India}
\\[1.4em]
\parbox[t]{0.32\textwidth}{\centering
  \textbf{Ayush Raj}\\
  Ashoka University, India} &
\parbox[t]{0.32\textwidth}{\centering
  \textbf{Saptarishi Dhanuka}\\
  Ashoka University, India} &
\parbox[t]{0.32\textwidth}{\centering
  \textbf{Parthasarathi Mukhopadhyay}\\
  Ashoka University, India}
\\[1.4em]
\multicolumn{3}{c}{%
  \parbox[t]{0.32\textwidth}{\centering
    \textbf{Sandeep Juneja}\\
    Ashoka University, India}}
\end{tabular}
}

\begin{document}
\maketitle

\begin{abstract}
Local heatwave hazard depends on fine-scale air temperature, but ground stations are sparse and reanalysis products such as ERA5 cannot resolve the terrain and land-surface contrasts that shape real heat exposure. We present a conditional diffusion emulator for high-resolution 2-m temperature downscaling, conditioned on static geography, a training climatology, exact-time ERA5 temperature, and solar and temporal features, guided at inference by score-based data assimilation (SDA): a differentiable Gaussian observation likelihood steers the diffusion score toward sparse revealed temperature observations without any retraining. On a controlled 32-case synthetic-grid protocol over AORC, guidance improves hidden-cell reconstruction over both ERA5 and a strong observation-proximal nearest-neighbor baseline once observation density reaches 1\% (RMSE 0.318 vs.\ 0.431~K, winning all 32 cases), while sparser regimes still favor direct interpolation. We further map the full guidance-strength landscape across three observation densities, showing that the optimal strength shifts systematically with density and that over-guiding causes sharp, predictable degradation -- giving a concrete operating recipe rather than a single untuned setting. The resulting fields are intended as a temperature layer for downstream heatwave-hazard products such as threshold exceedance and cumulative heat-burden. The present evidence is a controlled synthetic-grid validation; station-network and held-out-year evaluations are the next steps toward deployment.
\end{abstract}

\section{Introduction}

Fine-scale air temperature can support local heatwave preparedness and hazard assessment, but ground stations are sparse and reanalyses such as ERA5 cannot represent local terrain and land-surface contrasts \citep{hersbach2020era5}. Heat-health studies emphasize that a temperature field is a hazard input, and that absolute risk further depends on humidity, urban form, exposure, and social vulnerability \citep{kamath2023h3i}. The Analysis of Record for Calibration (AORC) is an hourly, 30-arc-second ($\approx$800~m) gridded weather record \citep{aorc2021}; we use it as a high-resolution reference target, not as independent station truth, so revealing synthetic observations from this already-interpolated analysis is a controlled proxy for, not a substitute for, assimilating real, noisier station reports. Diffusion models have separately been trained as kilometer-scale atmospheric downscaling and super-resolution systems, including residual corrective downscaling of forecasts \citep{mardani2025corrdiff} and foundation models that sharpen coarse forecasts into fine-scale multi-variable fields \citep{luitel2026aircast}, but without a mechanism for injecting sparse point observations at inference time.

The classical answer to combining a background field with sparse observations is data assimilation: optimal interpolation, 3D/4D-Var, and ensemble Kalman filtering all fuse a background prior with observations through an assumed, typically stationary, background-error covariance \citep{kalnay2003atmospheric}. We instead learn the background prior itself as a conditional diffusion model, whose score plays a role analogous to that covariance structure but is nonlinear, spatially adaptive, and conditioned on exact-time covariates rather than fixed a priori. A generative prior is deliberately preferred over a deterministic regression: sampling an ensemble yields a diagnostic uncertainty signal, and because the model targets the full conditional distribution rather than its mean, individual samples retain sharp fine-scale texture that a mean-squared-error regressor would smooth away -- the same score that produces this texture is exactly the object that observation guidance needs to steer at inference.

We propose a conditional diffusion emulator and approximate posterior guidance: the model maps static geography, a training climatology, exact-time ERA5 temperature, and solar and temporal features to an ensemble of high-resolution 2-m temperature anomaly fields, using the EDM (Elucidating the Design space of diffusion Models) parameterization \citep{karras2022edm} built on score-based generative modeling \citep{song2021scoresde}. At inference, a differentiable Gaussian observation likelihood modifies the diffusion score at revealed grid cells, following the broad idea of diffusion posterior sampling \citep{chung2023dps} and score-based data assimilation \citep{rozet2023sda}, in particular its recent application to sparse station observations in an unconditional diffusion weather prior at kilometer scale \citep{manshausen2024sda}. Unlike this unconditional setting, our prior is conditioned on exact-time ERA5 and static geography, and we characterize the guidance mechanism itself: sweeping eight guidance strengths across three observation densities shows that the optimal strength shifts systematically with density and that over-guiding produces a sharp, predictable failure mode -- a practical operating recipe rather than a single untuned setting. This separates reusable prior learning from event-specific assimilation and avoids retraining for each synthetic observation mask.

The paper focuses on the method and its controlled synthetic SDA evaluation. We establish the preprocessing, approximate-score construction, and sampling protocol, and compare against ERA5, a prior-only sampler, and nearest-observation filling across three observation densities. When 1\% of grid cells are randomly revealed, guidance reduces hidden-cell RMSE from 0.431~K (nearest-observation baseline) to 0.318~K, winning all 32 evaluation cases; at sparser densities, nearest-observation filling remains stronger, and Section~\ref{sec:heatwave} connects the resulting reconstruction layer to heatwave-hazard analysis. Station-network, geographic-transfer, and held-out-year evaluations are the next validation steps.

\section{Problem formulation and data}

Let $T(x,t)$ denote hourly AORC 2-m temperature at grid cell $x$. We train on Contiguous United States (CONUS) AORC from 2020--2021 and use 2022 only for validation and the reported synthetic evaluation. AORC is represented on its $\approx$800~m grid; 256$\times$256 patches are sampled with stride 128. Patches with more than 5\% invalid target pixels are excluded. Rather than modeling absolute temperature, we model an anomaly relative to a monthly--hourly, per-pixel training climatology $C(x,t)$, standardized by a global mean and standard deviation fit on the training years only; the exact transform and its inverse are given in Appendix~\ref{app:data}. Consequently, ERA5 is a conditioning field rather than the climatological baseline.

The 12 spatial conditioning channels are normalized topography, sky-view factor, climatological mean and standard deviation, exact-time ERA5 2-m temperature, cosine solar zenith angle, latitude, longitude, and sine/cosine encodings of day-of-year and hour-of-day. ERA5 is bilinearly interpolated onto the AORC grid. The noisy target is concatenated as a thirteenth input channel. All normalization statistics and the climatology are computed from the training years only.

\section{Conditional diffusion emulator}

\paragraph{Architecture and training.} We use a conditional U-Net based on the EDM parameterization, with base width 64, channel multipliers $(1,2,4,8)$, two residual blocks per resolution, and attention at 32$\times$32. The network has 69,324,125 trainable parameters. Training samples $\log\sigma\sim\mathcal{N}(-1.2,1.22^2)$ and uses the EDM denoising objective with $\sigma_{\mathrm{data}}=1$. We use AdamW with learning rate $1.5\times10^{-4}$, global batch size 24, gradient clipping at 1, an exponential moving average (EMA) of the weights with decay 0.999, bfloat16 mixed precision, and random seed 42, following training-dynamics practice for EDM-style networks \citep{karras2024edm2}. Training runs for a total of 250,000 steps; the SDA teacher is the EMA checkpoint at global step 205,000, a temporally smoothed snapshot of the weights taken near the end of this run rather than the raw step-205,000 weights.

\paragraph{Prior sampling.} For the observation-free conditional prior, we initialize with Gaussian noise and integrate 32 rho-spaced noise levels from $\sigma_{\max}=80$ to $\sigma_{\min}=0.02$ using the EDM Heun update ($\rho=7$). This produces an ensemble of plausible high-resolution anomaly fields conditioned on the same covariates.

\section{Score-based data assimilation}

Following the score-based data assimilation framework of \citet{rozet2023sda} and its adaptation to km-scale weather by \citet{manshausen2024sda}, who inject sparse weather-station observations into an unconditional diffusion prior via a guided score at inference time, we adapt this observation-guidance mechanism to our conditional temperature emulator. At each noise level, the trained denoiser induces a prior score over the anomaly field, exactly as in standard diffusion sampling. We additionally define a differentiable Gaussian observation likelihood that compares the denoiser's current estimate against whichever sparse, exact-time temperature values are revealed at a given inference call, masked to only those cells. Differentiating this likelihood through the denoiser and subtracting a weighted version of the resulting gradient from the prior score yields an approximate guided (posterior) score; substituting the corresponding guided denoiser estimate into the usual Euler update at each of the 32 noise levels turns the unconditional sampler into an observation-guided one, without any retraining of the network. The full construction, including the exact likelihood, the guidance weight, and the noise-level-dependent softening term, is given in Appendix~\ref{app:sda}.

Intuitively, the likelihood only ever evaluates the denoiser at the handful of revealed cells, while hidden cells are pulled toward consistency with those observations purely through the network's learned spatial correlations and its Jacobian -- this is what lets a small number of stations reshape a whole high-resolution patch. The guidance strength and the assumed observation error jointly control how hard the sampler is pulled toward the data versus the learned prior; the reported main results use a fixed, moderate guidance setting and an assumed physical observation error of 0.5~K, with the full sensitivity to these choices explored in Appendix~\ref{app:sda} and the guidance-strength sweep (Appendix~\ref{app:guidance-sweep}).

\section{Synthetic SDA downscaling evaluation}

\paragraph{Protocol.} We evaluate the step-205,000 EMA teacher on a fixed 32-case AORC 2022 validation manifest. Each patch uses an independently generated random valid-cell mask with seed 20250814. The 0.01\%, 0.1\%, and 1\% masks are nested subsets, with mean observed counts of 6.94, 67.06, and 655.25 cells per 256$\times$256 patch. This controlled synthetic-grid protocol isolates reconstruction from station siting and representativeness. Four guided members are sampled in one batch with 32 Euler steps.

\paragraph{Metrics and baselines.} We compute RMSE and MAE in kelvin only over valid hidden cells, excluding revealed cells, and report macro-means over the 32 per-patch scores. ERA5 is the exact-time 2-m temperature at each hidden cell. The prior-only baseline uses the same Euler-guided sampler with $\lambda_{\mathrm{obs}}=0$; it is distinct from the Heun model-only diagnostic in the appendix. The nearest baseline copies the nearest revealed normalized AORC anomaly and reconstructs with the target cell's local climatology, giving ``nearest observed anomaly + local climatology,'' a strong correlation-preserving comparator rather than raw nearest-temperature filling. We additionally run two classical spatial interpolators directly on the physical temperature field rather than the anomaly: modified Shepard inverse-distance-weighted interpolation \citep{shepard1968idw}, reconstructing each hidden cell from its eight nearest revealed cells with inverse-square-distance weights, and local ordinary kriging over the same eight-neighbor window, using a per-patch exponential variogram fit from the revealed cells.

\section{Results}

\begin{table}[t]
\caption{Synthetic sparse-observation downscaling on 32 AORC 2022 patches. Values are hidden-cell RMSE/MAE in K; macro means over patches. SDA uses the 205k EMA checkpoint, four members, 32 Euler steps, and assumed observation error 0.5~K. Shepard and Kriging are eight-neighbor classical spatial interpolators applied to the raw temperature field.}
\label{tab:sda}
\centering
\scriptsize
\begin{tabular}{lcccccc}
\toprule
Observed & ERA5 & Prior-only & Nearest+clim. & Shepard & Kriging & SDA \\
\midrule
0.01\% & 1.931/1.584 & 2.546/2.288 & \textbf{1.486/1.111} & 2.096/1.617 & 2.165/1.643 & 2.327/2.069 \\
0.1\%  & 1.931/1.584 & 2.546/2.288 & \textbf{0.774/0.531} & 1.125/0.797 & 1.081/0.765 & 1.086/0.875 \\
1\%    & 1.931/1.584 & 2.546/2.288 & 0.431/0.272 & 0.626/0.408 & 0.593/0.394 & \textbf{0.318/0.230} \\
\bottomrule
\end{tabular}
\end{table}

\begin{figure}[t]
\centering
\includegraphics[width=0.94\linewidth]{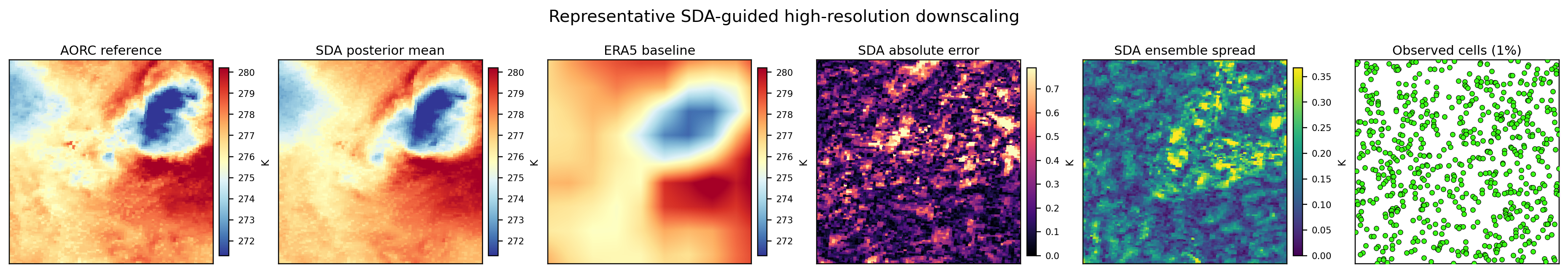}
\caption{Representative 1\% SDA downscaling case from the frozen 205k-checkpoint evaluation. The posterior mean follows the AORC field while sparse observations steer the learned prior; ensemble spread highlights spatially uncertain regions.}
\label{fig:sda-case}
\end{figure}

At 1\% observations, SDA reduces RMSE from 0.431 to 0.318~K relative to nearest observed anomaly plus local climatology, a 0.1125~K macro-mean improvement, and is far below ERA5 at 1.931~K. A paired nonparametric bootstrap over the 32 patch-level RMSE differences gives a 95\% interval of 0.076--0.156~K; SDA wins RMSE on all 32 cases. This is a demanding comparison: nearest filling directly propagates a revealed AORC anomaly and therefore preserves the local signal wherever nearby cells are correlated. SDA's consistent improvement shows that the learned spatial prior contributes useful sub-patch structure beyond pointwise propagation. Density reveals complementary operating regimes: nearest filling is strongest at 0.01\% and 0.1\%, whereas SDA is strongest at 1\%. The prior-only sampler establishes that the gain comes from combining the learned prior with observations.

\section{From guided temperature fields to heatwave-risk analysis}
\label{sec:heatwave}

The central result is a practical route from sparse measurements to a fine-scale
physical heat-hazard field. At 1\% observations, SDA beats the strong
observation-proximal nearest baseline on all 32 validation patches, reducing
hidden-cell RMSE by 0.112~K. This matters for heatwave analysis because
threshold exceedance and duration can vary over distances much smaller than a
reanalysis grid cell. A guided field can therefore expose spatial contrasts
that pointwise observation propagation and coarse reanalysis smooth away.

During an extreme-heat episode, the reconstructed 2-m field can provide the
temperature layer for maps of threshold exceedance or cumulative degree-hours.
These maps can then be joined with humidity, urban morphology, population
exposure, and social vulnerability, following the modular heat-health framing
of \citet{kamath2023h3i}. The present work establishes the reconstruction
component; event-based heatwave and health-risk studies are a direct next step.

The evaluation is intentionally controlled: the 32 cases use a 2022 synthetic
grid-observation protocol and the checkpoint was selected using that period's
validation loss. Four-member spread is consequently a diagnostic uncertainty
signal, not a calibrated tail probability. Station representativeness,
held-out-year transfer, humidity-aware hazard metrics, and threshold outcomes
should be evaluated before operational deployment. These are extensions of the
validated physical layer rather than changes to the SDA mechanism.

\section{Conclusion}

We presented a conditional diffusion emulator and denoiser-guided approximate
posterior method for sparse-grid temperature downscaling. The method injects
observations through a denoiser-differentiated likelihood without retraining.
On the fixed 32-case AORC 2022 protocol, SDA improves over ERA5 and a strong
nearest-observation baseline at 1\% revealed cells, winning every patch and
reducing hidden-cell RMSE by 0.112~K. This demonstrates that the learned
spatial prior can reconstruct useful fine-scale structure beyond direct
pointwise propagation. The resulting fields offer a concrete temperature
layer for heatwave threshold, duration, and exposure analyses. Future work
should calibrate guidance to density and observation noise, evaluate held-out
events with station observations and humidity, and connect reconstructed
hazard fields to heat-health outcomes.

\begin{ack}
The authors thank the open AORC and ERA5 data communities. All reported metrics and figures are generated from the experiment artifacts accompanying this manuscript.
\end{ack}

\clearpage
\appendix
\section{Extended methodology and diagnostics}

This appendix records the complete mathematical specification and auxiliary
development diagnostics. The diagnostics use the checkpoint and protocol stated
in each caption and are not pooled as a single benchmark: the main synthetic
SDA table uses the step-205k teacher, the guidance stress test uses a separate
step-165k EMA teacher, and the model-only diagnostic uses a 250k checkpoint.

\subsection{Data, target transform, and conditioning}
\label{app:data}

Training uses CONUS AORC hourly 2-m temperature from 2020--2021 and evaluates
on 2022. A per-pixel monthly--hourly climatology $C(x,t)$ is computed from the
training years only. With $T(x,t)$ denoting AORC temperature, the normalized
anomaly target is
\begin{equation}
 z(x,t)=\frac{(T(x,t)-C(x,t))-\mu_a}{s_a},
 \qquad \mu_a=0.7182066~\mathrm{K},\quad s_a=4.7507637~\mathrm{K}.
 \label{eq:app-target}
\end{equation}
The physical downscaled field is therefore
\begin{equation}
 \widehat{T}(x,t)=C(x,t)+s_a\widehat{z}(x,t)+\mu_a.
 \label{eq:app-inverse}
\end{equation}
Each $256\times256$ patch supplies twelve conditioning channels: normalized
topography, sky-view factor, climatological mean and standard deviation,
exact-time ERA5 temperature, cosine solar zenith angle, normalized latitude and
longitude, and sine/cosine encodings of day-of-year and hour-of-day. The noisy
target is concatenated as a thirteenth model input channel. Invalid target
pixels are masked in every loss and metric.

\subsection{EDM preconditioning, objective, and sampling}
\label{app:edm}

For clean normalized target $x_0$ and Gaussian noise $\epsilon\sim
\mathcal{N}(0,I)$, training draws
\begin{equation}
 \log\sigma\sim\mathcal{N}(P_{\mathrm{mean}},P_{\mathrm{std}}^2),
 \qquad (P_{\mathrm{mean}},P_{\mathrm{std}})=(-1.2,1.22),
 \qquad x_\sigma=x_0+\sigma\epsilon.
 \label{eq:app-noise}
\end{equation}
The EDM preconditioner uses $\sigma_{\mathrm{data}}=1$:
\begin{align}
 c_{\mathrm{skip}}(\sigma)&=\frac{\sigma_{\mathrm{data}}^2}{\sigma^2+\sigma_{\mathrm{data}}^2}, &
 c_{\mathrm{out}}(\sigma)&=\frac{\sigma\sigma_{\mathrm{data}}}{\sqrt{\sigma^2+\sigma_{\mathrm{data}}^2}},\\
 c_{\mathrm{in}}(\sigma)&=\frac{1}{\sqrt{\sigma^2+\sigma_{\mathrm{data}}^2}}, &
 c_{\mathrm{noise}}(\sigma)&=\frac{\log\sigma}{4}.
 \label{eq:app-precond}
\end{align}
If $F_\theta$ is the U-Net output, the denoised prediction is
\begin{equation}
 D_\theta(x_\sigma,\sigma,c)=c_{\mathrm{skip}}x_\sigma+
 c_{\mathrm{out}}F_\theta\!\left(c_{\mathrm{in}}x_\sigma,c_{\mathrm{noise}}(\sigma),c\right).
 \label{eq:app-denoiser}
\end{equation}
Here the unsubscripted $c$ is the 13-channel conditioning stack from Appendix~\ref{app:data} (the 12 static and exact-time covariates plus the noisy target), not to be confused with the noise-dependent scalar preconditioning coefficients $c_{\mathrm{skip}}$, $c_{\mathrm{out}}$, $c_{\mathrm{in}}$, $c_{\mathrm{noise}}$ defined above. With valid-pixel indicator $V_i\in\{0,1\}$, equal to 1 where cell $i$ has a finite (non-missing) AORC record and 0 at fill-value cells such as data gaps, the implemented weighted denoising objective is
\begin{equation}
 \mathcal{L}_{\mathrm{EDM}}=\mathbb{E}_{x_0,\sigma,\epsilon}
 \left[\frac{\sum_i V_i\,w(\sigma)\big(D_{\theta,i}-x_{0,i}\big)^2}
 {\max(1,\sum_iV_i)}\right],
 \qquad
 w(\sigma)=\frac{\sigma^2+\sigma_{\mathrm{data}}^2}
 {\big(\sigma\sigma_{\mathrm{data}}\big)^2}.
 \label{eq:app-edm-loss}
\end{equation}

The model has base width 64, channel multipliers $(1,2,4,8)$, two residual
blocks per resolution, attention at $32\times32$, and 69,324,125 trainable
parameters. AdamW uses learning rate $1.5\times10^{-4}$, zero weight decay,
global batch size 24, gradient clipping at 1, EMA decay 0.999, bfloat16 mixed
precision, and seed 42.

The deterministic prior sampler uses $N=32$ noise levels with $\sigma_{\max}=80$,
$\sigma_{\min}=0.02$, and $\rho=7$:
\begin{equation}
 \sigma_i=\left(\sigma_{\max}^{1/\rho}+\frac{i}{N-1}
 \left(\sigma_{\min}^{1/\rho}-\sigma_{\max}^{1/\rho}\right)\right)^\rho,
 \qquad i=0,\ldots,N-1.
 \label{eq:app-schedule}
\end{equation}
For the unconditional prior, $d_i=(x_i-D_i)/\sigma_i$ and the Euler proposal
is $x_{i+1}^{\mathrm{E}}=x_i+(\sigma_{i+1}-\sigma_i)d_i$. Except at the final
step, a second denoiser evaluation gives $d_{i+1}$ and the Heun update
\begin{equation}
 x_{i+1}=x_i+\frac{\sigma_{i+1}-\sigma_i}{2}(d_i+d_{i+1}).
 \label{eq:app-heun}
\end{equation}

\subsection{Score-based data assimilation}
\label{app:sda}

At each noise level, the denoiser induces the prior score
\begin{equation}
 s_{\mathrm{prior}}(x_\sigma)=\frac{D_\theta(x_\sigma,\sigma,c)-x_\sigma}{\sigma^2}.
 \label{eq:app-prior-score}
\end{equation}
Let $M_i\in\{0,1\}$ indicate revealed cells and let $y_i$ be the normalized
observations. The differentiable Gaussian likelihood is
\begin{equation}
 \mathcal{L}_{\mathrm{obs}}(x_\sigma)=\frac{1}{2}\sum_i
 \frac{\left[M_i\left(D_{\theta,i}-y_i\right)\right]^2}
 {\sigma_{\mathrm{obs}}^2+\gamma\sigma^2},
 \qquad
 s_{\mathrm{obs}}=-\nabla_{x_\sigma}\mathcal{L}_{\mathrm{obs}}.
 \label{eq:app-obs-loss}
\end{equation}
The $\gamma\sigma^2$ term inflates the effective observation variance in proportion to the current diffusion noise level, so the likelihood constrains the sampler weakly at the large $\sigma$ used early in sampling (where $D_\theta$ is still a coarse estimate) and progressively more strongly as $\sigma\to\sigma_{\mathrm{min}}$ and the denoiser output sharpens.
The posterior-guided score and corresponding clean estimate are
\begin{align}
 s_{\mathrm{post}}&=s_{\mathrm{prior}}+\lambda_{\mathrm{obs}}s_{\mathrm{obs}},\\
 D_{\mathrm{post}}&=x_\sigma+\sigma^2s_{\mathrm{post}}.
 \label{eq:app-post-score}
\end{align}
SDA uses the Euler proposal $x_{i+1}=x_i+(\sigma_{i+1}-\sigma_i)
(x_i-D_{\mathrm{post},i})/\sigma_i$ at all 32 noise levels. The production
configuration uses $\gamma=0.001$, sum reduction over observed cells,
$\lambda_{\mathrm{obs}}=1$, no score clipping (the guided score $s_{\mathrm{post}}$
is used as computed, without clamping its magnitude before the Euler update),
and an assumed physical observation error of $0.5$~K, corresponding to
$\sigma_{\mathrm{obs}}=0.5/s_a=0.1052462$ in normalized units. Synthetic
observations contain no added noise.

For hidden-valid index set
\begin{equation}
 \mathcal{H}=\{i:V_i(1-M_i)=1\},
 \label{eq:app-hidden}
\end{equation}
the reported per-case errors are
\begin{equation}
 \mathrm{RMSE}=\sqrt{\frac{1}{|\mathcal{H}|}\sum_{i\in\mathcal{H}}
 (\widehat{T}_i-T_i)^2},
 \qquad
 \mathrm{MAE}=\frac{1}{|\mathcal{H}|}\sum_{i\in\mathcal{H}}
 |\widehat{T}_i-T_i|.
 \label{eq:app-metrics}
\end{equation}
The ensemble 90\% coverage is the fraction of hidden cells containing the
truth between the empirical 5th and 95th percentiles. For ensemble CDF $F$ and
truth $y$, the continuous ranked probability score is
\begin{equation}
 \mathrm{CRPS}(F,y)=\mathbb{E}_{X\sim F}|X-y|
 -\frac{1}{2}\mathbb{E}_{X,X'\sim F}|X-X'|.
 \label{eq:app-crps}
\end{equation}

\section{Model-only emulator diagnostics}

The selected model-only evaluation uses the 250k EMA checkpoint, four fixed
AORC 2022 patches, 16 ensemble members, and 32 Heun steps. It is a development
diagnostic, not a baseline for the 205k guided evaluation. The macro means are:

\begin{table}[t]
\caption{Four-patch model-only spatial evaluation at the 250k checkpoint.}
\centering
\small
\begin{tabular}{lrr}
\toprule
Method & MAE (K) & RMSE (K) \\
\midrule
Diffusion ensemble mean & 1.736 & 1.909 \\
ERA5 T2m & 1.230 & 1.559 \\
Training climatology & 4.581 & 4.731 \\
\bottomrule
\end{tabular}
\end{table}

The spatial output confirms that the emulator recovers broad temperature
structure and improves substantially over climatology, but its ensemble mean
is smoother than AORC.

\begin{figure}[t]
\centering
\includegraphics[width=0.96\linewidth]{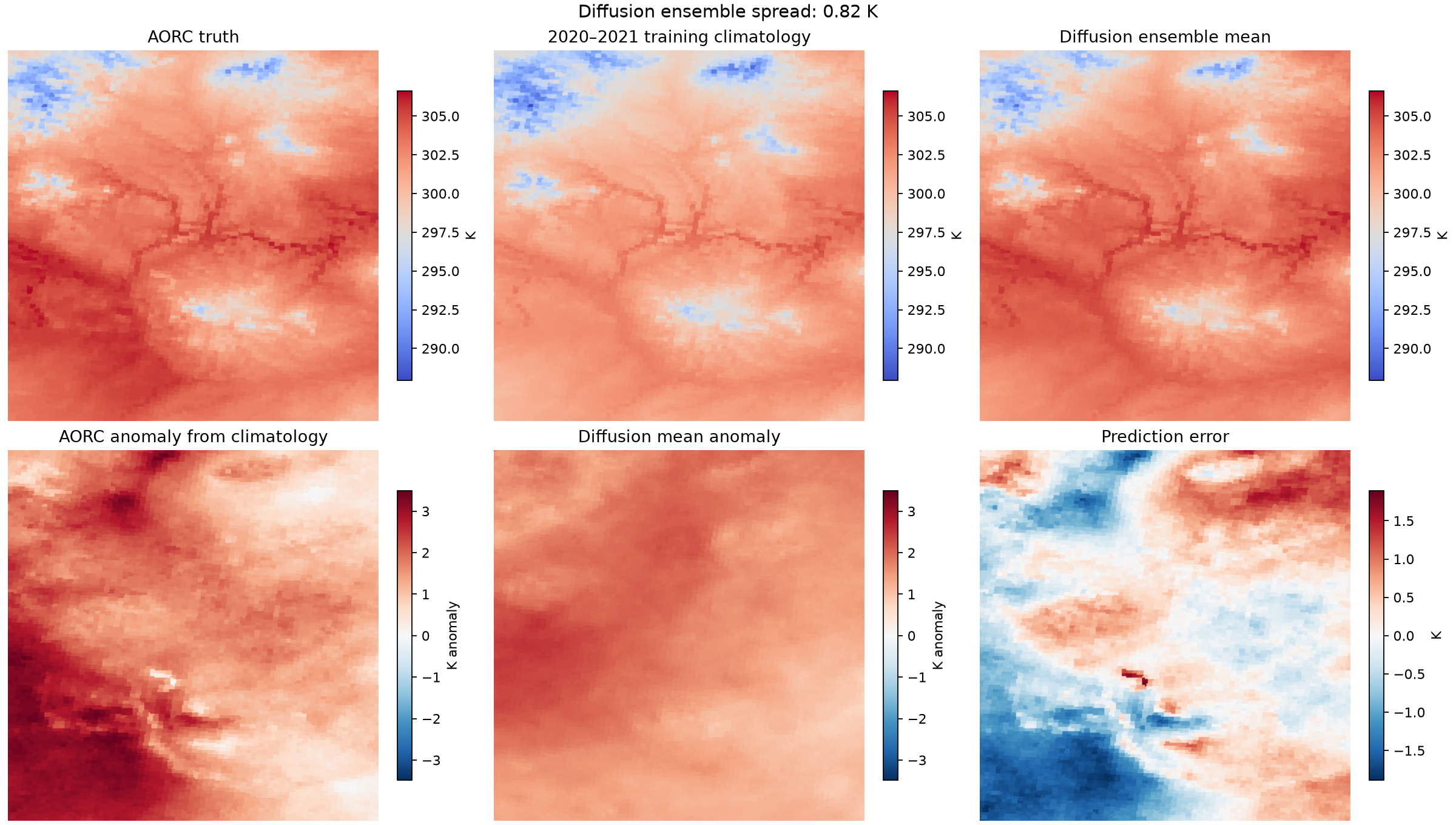}
\caption{Representative 250k-checkpoint emulator output. The diffusion mean follows the broad AORC temperature structure while smoothing fine-scale anomaly detail; the ensemble spread is 0.82~K for this patch.}
\label{fig:app-model-spatial}
\end{figure}

\section{Guidance-strength sweep}
\label{app:guidance-sweep}

The completed guidance stress test uses the step-165k EMA checkpoint, ten matched
2022 validation cases per combination, eight posterior members, 64 guided
steps, exact synthetic observations, and $\lambda_{\mathrm{obs}}\in
\{0.5,1,2,5,8,10,20,50\}$. It covers 1\%, 5\%, and 10\% observed cells. A
matched $\lambda=0$ run is the prior-only reference for each density. Because
the observation likelihood is sum-reduced, increasing either $\lambda$ or the
number of observed cells increases the aggregate observation force. These runs
were not used to tune the step-205k main result, because their checkpoint,
sampler length, and ensemble size differ.

\begin{table}[t]
\caption{Guidance sweep hidden-cell RMSE/MAE (K), averaged over ten matched cases. The eight columns are the nominal observation-guidance strengths.}
\centering
\scriptsize
\resizebox{\linewidth}{!}{%
\begin{tabular}{lrrrrrrrr}
\toprule
Observed & $\lambda=.5$ & $1$ & $2$ & $5$ & $8$ & $10$ & $20$ & $50$ \\
\midrule
1\%  & .633/.480 & .541/.408 & .478/.357 & \textbf{.414/.304} & 1.905/1.371 & 3.115/2.319 & 5.844/4.481 & 106.233/14.718 \\
5\%  & .439/.330 & \textbf{.381/.284} & 1.343/1.021 & 3.070/2.317 & 4.086/3.079 & 4.726/3.605 & 7.489/5.820 & $6.279\times10^3/1.452\times10^3$ \\
10\% & \textbf{.377/.282} & 1.149/.888 & 1.986/1.504 & 3.339/2.527 & 4.538/3.446 & 5.182/3.980 & 10.090/7.853 & $1.338\times10^5/1.159\times10^4$ \\
\bottomrule
\end{tabular}}
\end{table}

The best strength shifts downward as observations become denser: $\lambda=5$
at 1\%, $\lambda=1$ at 5\%, and $\lambda=0.5$ at 10\%. At 1\%, the best
SDA reaches $0.414\pm0.123$~K RMSE and $0.304\pm0.086$~K MAE, versus
$0.530\pm0.189$/$0.333\pm0.110$ for nearest-observation filling,
$0.853\pm0.542$/$0.565\pm0.362$ for Shepard interpolation,
$0.836\pm0.565$/$0.570\pm0.404$ for local ordinary kriging, and
$2.238\pm0.996$/$1.850\pm0.857$ for ERA5. At 5\%, best SDA is
$0.381\pm0.097$/$0.284\pm0.069$ versus nearest
$0.361\pm0.130$/$0.201\pm0.068$, Shepard
$0.587\pm0.389$/$0.370\pm0.249$, and kriging
$0.593\pm0.420$/$0.391\pm0.293$; at 10\%, best SDA is
$0.377\pm0.095$/$0.282\pm0.067$ versus nearest
$0.310\pm0.110$/$0.156\pm0.054$, Shepard
$0.504\pm0.339$/$0.313\pm0.215$, and kriging
$0.515\pm0.373$/$0.334\pm0.257$. Thus the sweep identifies a useful
intermediate-guidance regime rather than supporting arbitrarily large
guidance.

\begin{figure}[t]
\centering
\includegraphics[width=0.49\linewidth]{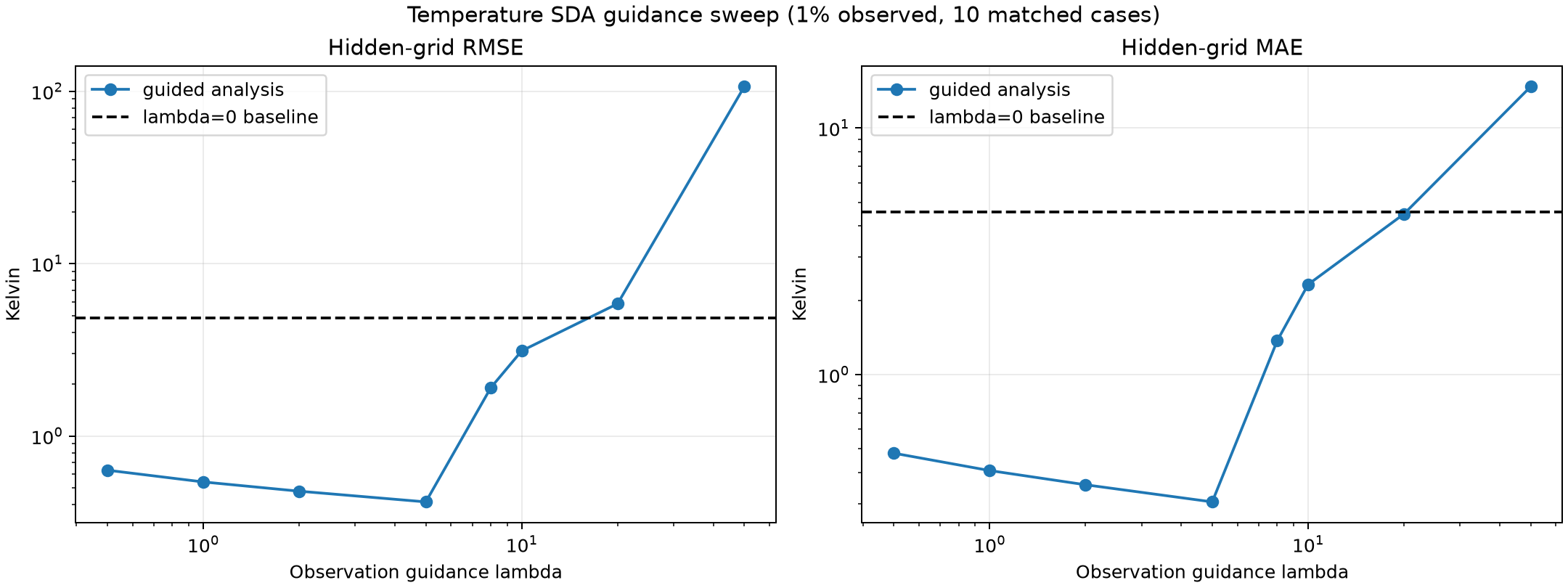}\hfill
\includegraphics[width=0.49\linewidth]{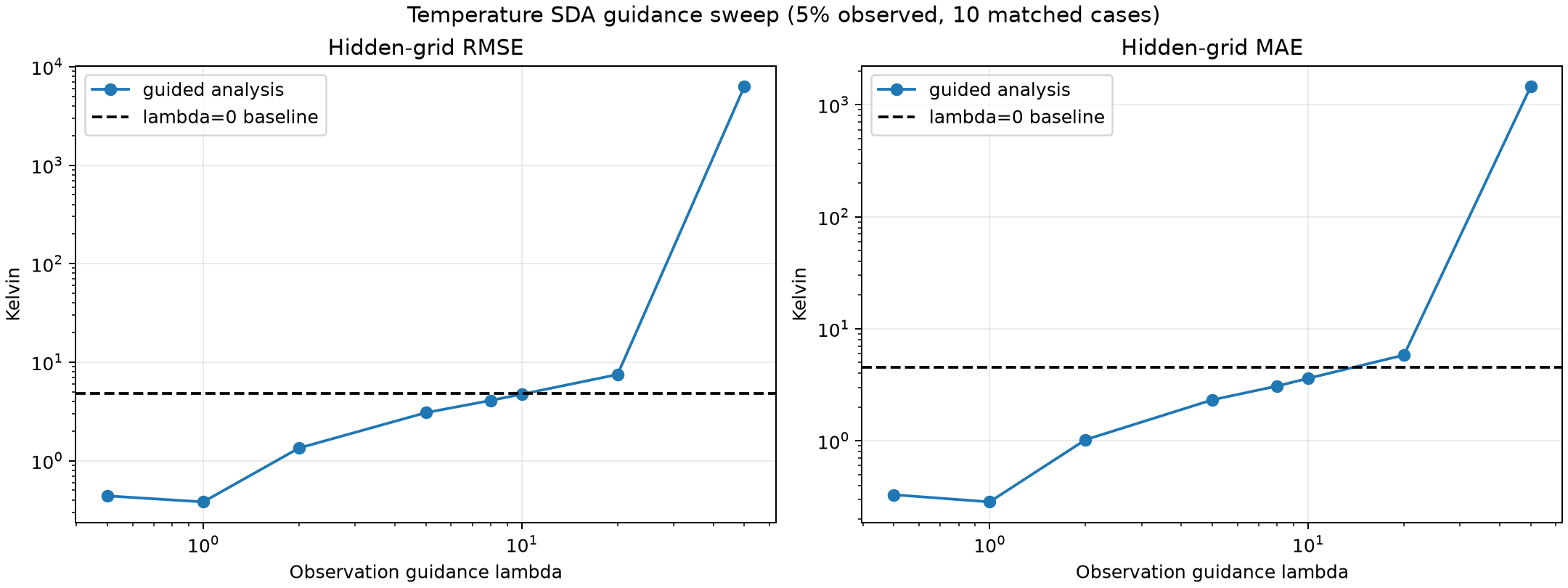}
\caption{Guidance-strength sweeps for 1\% and 5\% observed cells. Both axes are logarithmic; the dashed line is the matched $\lambda=0$ baseline. Errors decrease initially, then rise sharply as the observation force overwhelms the learned prior.}
\label{fig:app-guidance-low}
\end{figure}

\begin{figure}[t]
\centering
\includegraphics[width=0.96\linewidth]{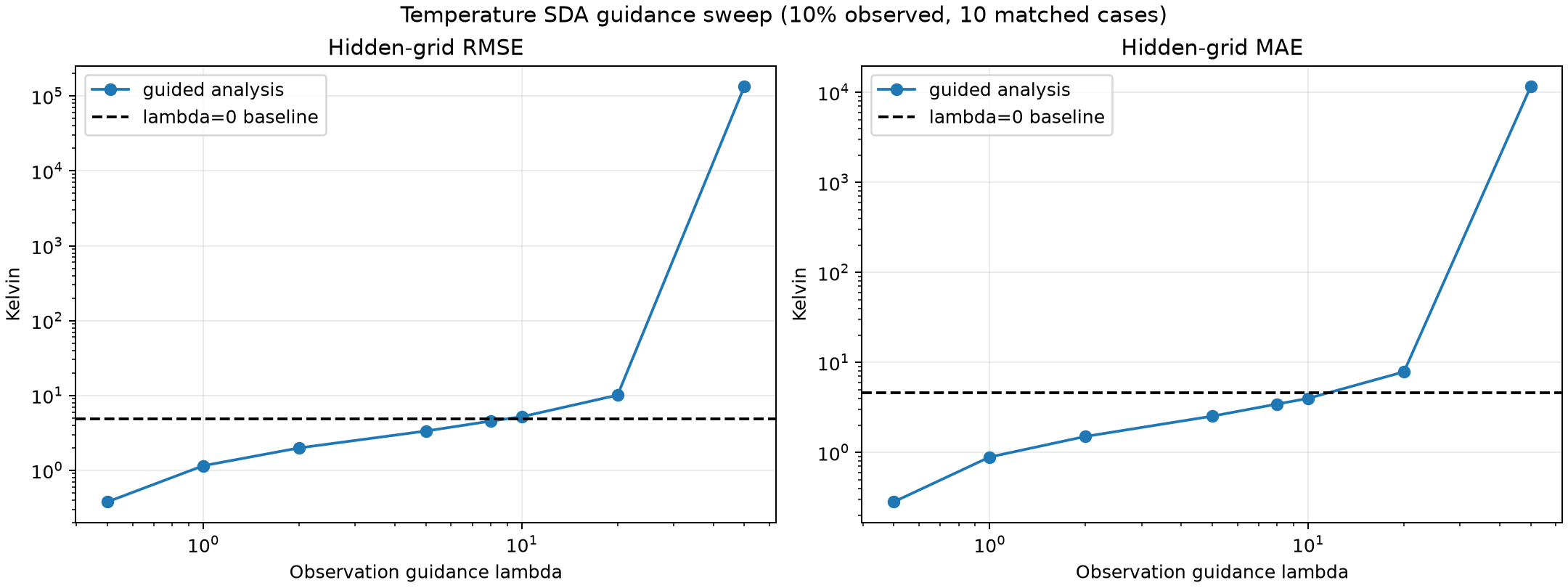}
\caption{Guidance-strength sweep for 10\% observed cells. The best setting is $\lambda=0.5$; stronger guidance rapidly degrades hidden-cell performance.}
\label{fig:app-guidance-10}
\end{figure}

\begin{figure}[t]
\centering
\includegraphics[width=0.96\linewidth]{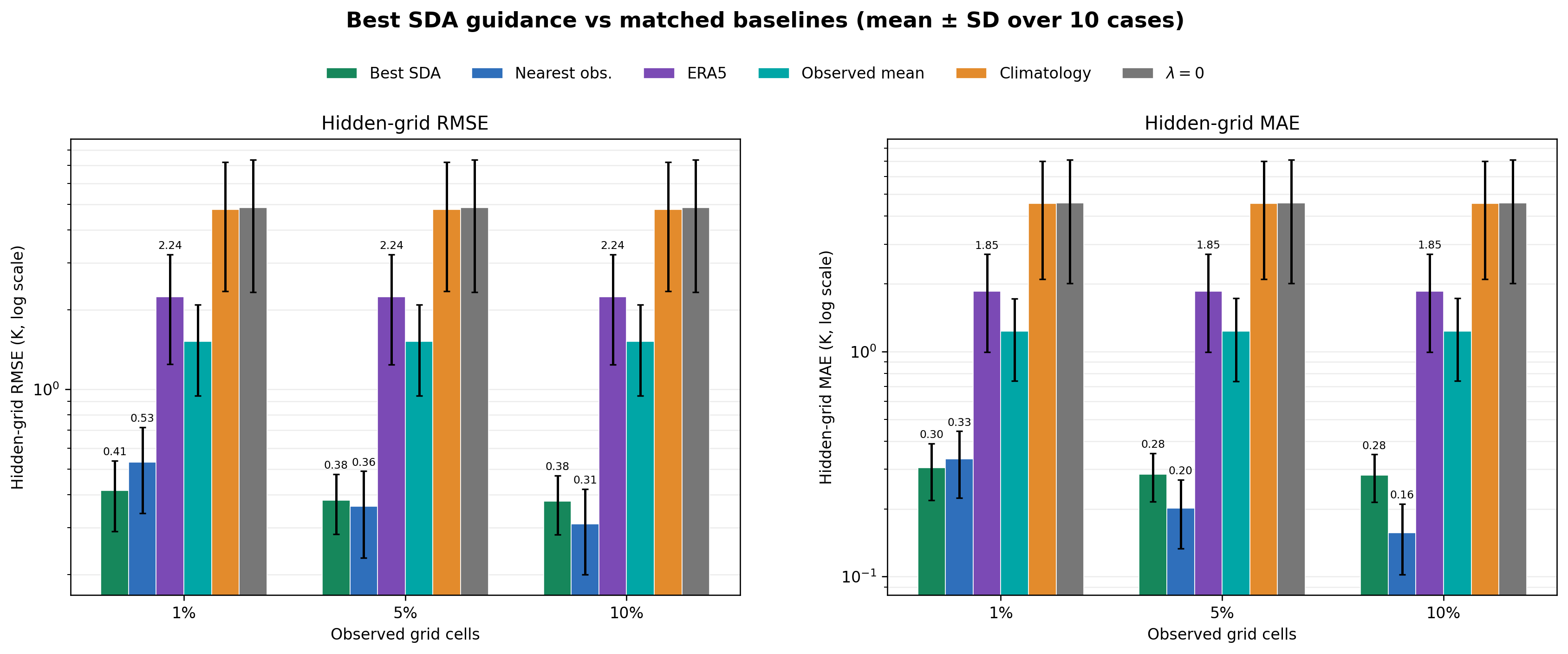}
\caption{Best guidance at each density versus matched baselines over ten cases. Error bars show sample standard deviation; logarithmic axes retain both sub-Kelvin SDA errors and unstable high-guidance values.}
\label{fig:app-guidance-baselines}
\end{figure}

\begin{figure}[t]
\centering
\includegraphics[width=0.96\linewidth]{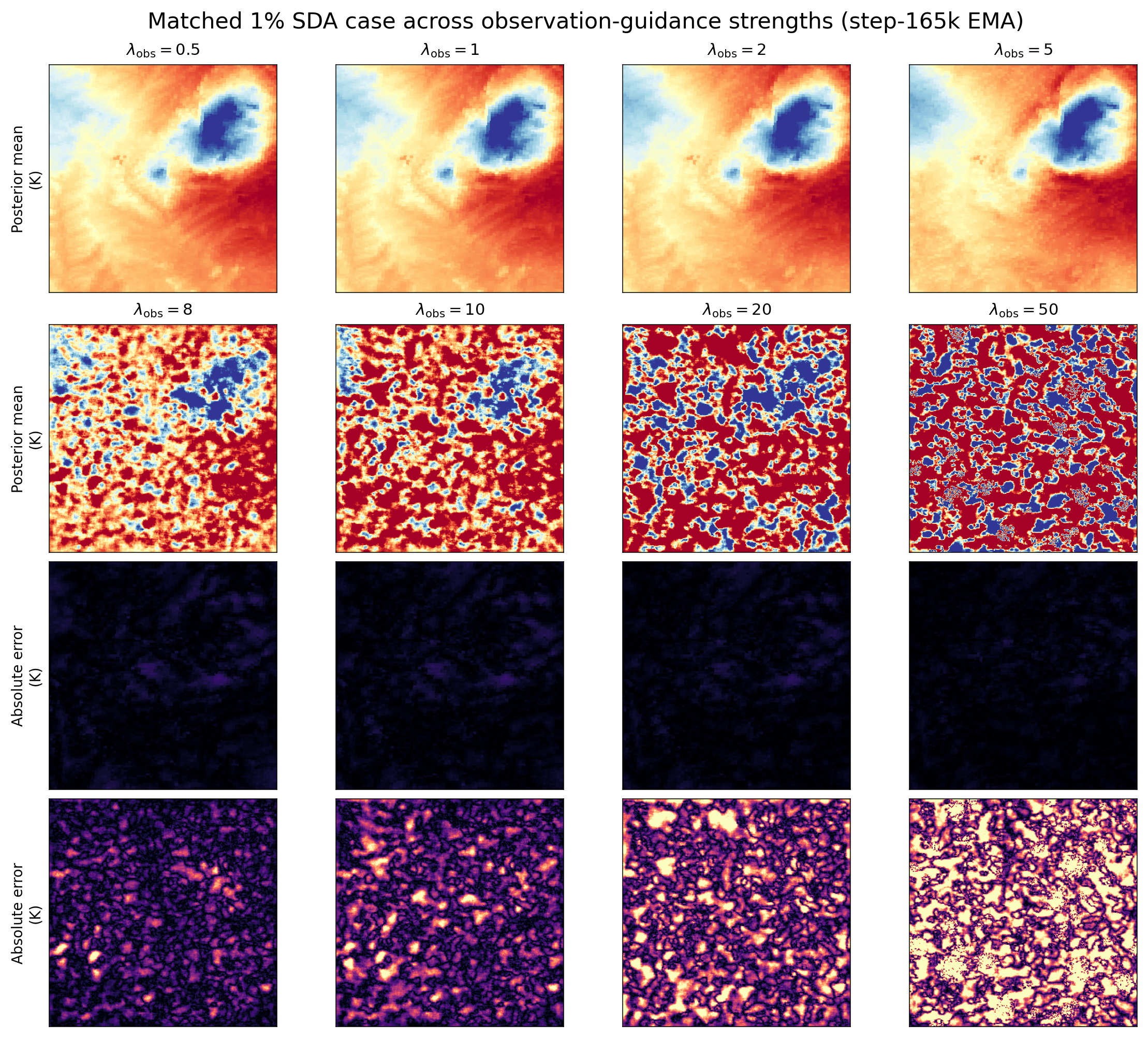}
\caption{One matched 1\% case across guidance strengths. Rows 1--2 show posterior means (display clipped to the truth range); rows 3--4 show absolute errors on a common robust scale. Spatial artifacts become conspicuous from $\lambda=8$ onward and are extreme at $\lambda=50$.}
\label{fig:app-guidance-spatial}
\end{figure}

The qualitative trend is consistent across densities: low-to-moderate guidance
reduces hidden-cell error, while high guidance produces spatially localized
speckle, guide-point overfitting, and eventually numerical instability. The
effect is operationally important: at 1\% observed cells, mean hidden RMSE
rises from 0.414~K at $\lambda=5$ to 1.905~K at $\lambda=8$, 5.844~K at
$\lambda=20$, and 106.233~K at $\lambda=50$. At 10\%, the corresponding
values are 0.377, 4.538, 10.090, and $1.338\times10^5$~K. The observation
fit can improve while hidden-cell performance collapses, which is why guidance
must be tuned jointly with observation density and likelihood scaling.

\section{From downscaled temperature to heatwave-hazard layers}

The model output is a physical temperature field that can serve as the
temperature layer in a heat-health assessment. For an episode indexed by hours $t$ and a
stakeholder-defined threshold $\tau$, a simple exceedance layer is
\begin{equation}
 H_\tau(x,t)=\mathbb{1}\!\left[\widehat{T}(x,t)\geq\tau\right],
 \qquad
 B_\tau(x)=\sum_{t\in\mathcal{E}}\Delta t\,
 \max\!\left(0,\widehat{T}(x,t)-\tau\right),
 \label{eq:heatwave-layer}
\end{equation}
where $\mathcal{E}$ is the episode window and $B_\tau$ is a cumulative
degree-hour burden. For an ensemble, the corresponding exceedance probability
is estimated by
\begin{equation}
 \widehat{p}_\tau(x,t)=\frac{1}{K}\sum_{k=1}^{K}
 \mathbb{1}\!\left[\widehat{T}^{(k)}(x,t)\geq\tau\right].
 \label{eq:heatwave-prob}
\end{equation}
These products show how sparse observations can steer the spatial distribution
and duration of an extreme-heat signal. They can subsequently be combined with
humidity, urban morphology, population exposure, and social vulnerability, as
emphasized by the H3I framework \citep{kamath2023h3i}; evaluating threshold skill
and health outcomes is a natural next stage.

\section{Additional synthetic SDA outputs}

The main paper shows one representative 1\% case. The following matched case
uses the step-205k teacher and illustrates how the posterior mean, error, and
ensemble spread change as the mask density increases.

\begin{figure}[t]
\centering
\includegraphics[width=0.96\linewidth]{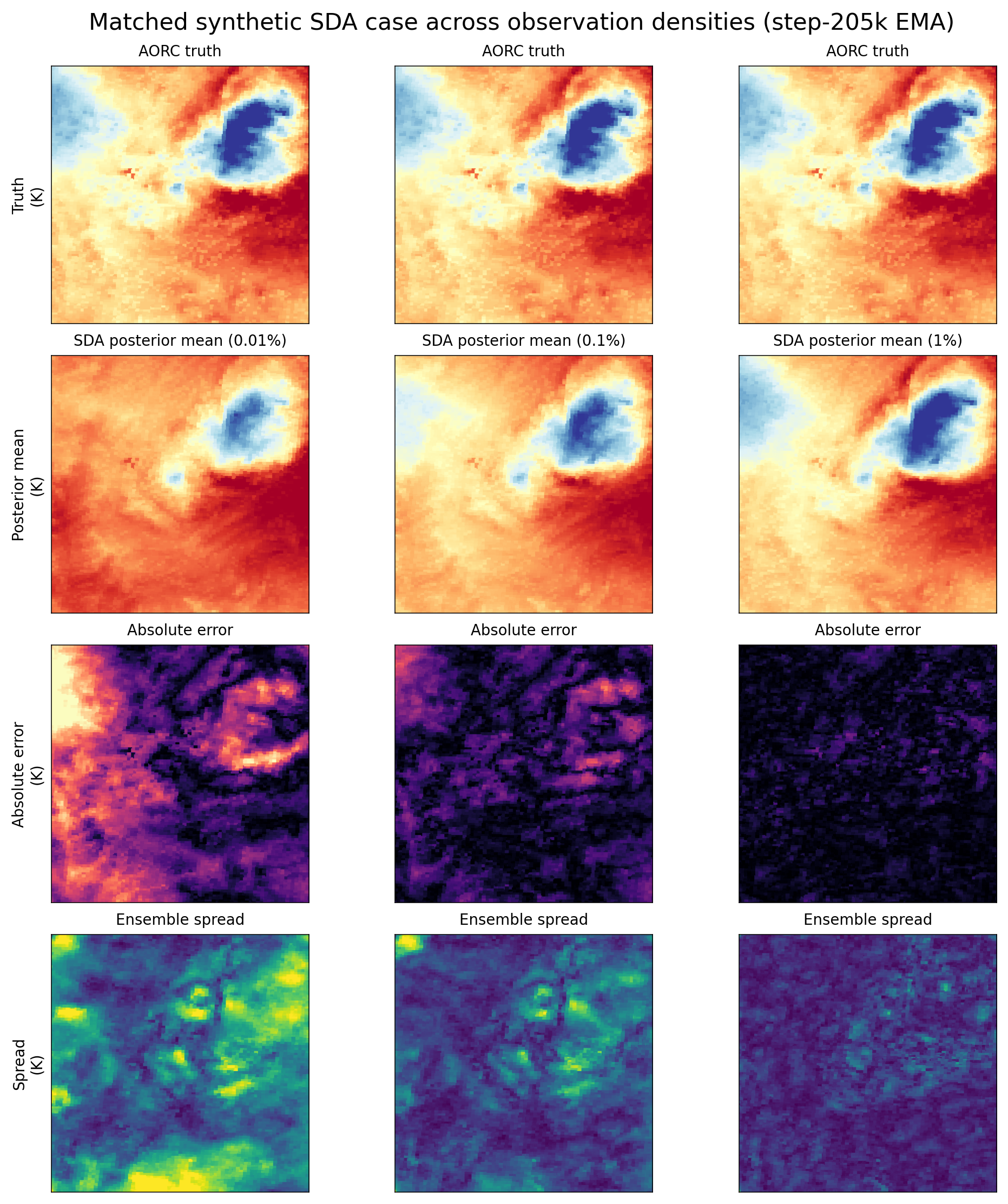}
\caption{Matched synthetic SDA case at 0.01\%, 0.1\%, and 1\% observed cells. More observations recover progressively finer structure and reduce hidden-cell error, while the ensemble spread contracts in the better-constrained regions.}
\label{fig:app-sda-density}
\end{figure}

The primary 32-case step-205k macro means are repeated here for reference:

\begin{table}[t]
\caption{Primary synthetic SDA macro means over 32 AORC 2022 patches. Values are RMSE/MAE in K.}
\centering
\scriptsize
\begin{tabular}{lrrrrrr}
\toprule
Observed & ERA5 & Prior & Nearest & Shepard & Kriging & SDA \\
\midrule
0.01\% & 1.931/1.584 & 2.546/2.288 & 1.486/1.111 & 2.096/1.617 & 2.165/1.643 & 2.327/2.069 \\
0.1\%  & 1.931/1.584 & 2.546/2.288 & 0.774/0.531 & 1.125/0.797 & 1.081/0.765 & 1.086/0.875 \\
1\%    & 1.931/1.584 & 2.546/2.288 & 0.431/0.272 & 0.626/0.408 & 0.593/0.394 & \textbf{0.318/0.230} \\
\bottomrule
\end{tabular}
\end{table}

At 1\%, SDA improves on nearest filling for all 32 cases under the frozen
protocol. The lower-density rows show the complementary regime in which
nearest observed anomaly plus local climatology is stronger. This density
dependence is also consistent with the guidance sweep.

\section{Follow-up robustness checks}

We reran the 1\% protocol with the same fixed 205k teacher, 32 cases, four
members, and 32 Euler steps, but without changing the frozen primary table.
Two fresh random masks, a regular approximately 10-cell-spacing grid, and
independent Gaussian perturbations of standard deviation 0.5~K were evaluated.
These extend the 2022 validation evidence; station and blind-year tests are
the next step toward deployment.

\begin{table}[t]
\caption{Follow-up 1\% synthetic robustness checks. Values are hidden-cell
macro RMSE/MAE in K over 32 cases. ``Wins'' is the number of per-case SDA RMSE
wins against nearest observed anomaly plus local climatology.}
\centering
\scriptsize
\begin{tabular}{lrrr}
\toprule
Condition & SDA & Nearest & Wins \\
\midrule
Random seed 20250815 & .321/.230 & .433/.273 & 31/32 \\
Random seed 20250816 & .318/.229 & .435/.274 & 32/32 \\
Regular grid & .288/.211 & .374/.235 & 27/32 \\
Random + 0.5~K noise & .360/.268 & .679/.516 & 32/32 \\
\bottomrule
\end{tabular}
\end{table}

The random-mask gain is stable across the two new seeds. The regular grid
improves both methods but retains an SDA macro advantage. Adding 0.5~K
observation noise degrades SDA relative to noiseless observations, yet SDA
remains better than nearest filling under this observation-noise perturbation.
Next evaluations should address station representativeness, density-aware
guidance, and model selection on genuinely held-out years.


\begin{thebibliography}{9}

\bibitem[Fall et~al.(2023)Fall, Kitzmiller, Pavlovic, Zhang, Patrick, St.
Laurent, Trypaluk, Wu, and Miller]{aorc2021}
G. Fall, D. Kitzmiller, S. Pavlovic, Z. Zhang, N. Patrick, M.~St. Laurent, C.
Trypaluk, W. Wu, and D. Miller.
\newblock The {O}ffice of {W}ater {P}rediction's {A}nalysis of {R}ecord for
{C}alibration, version 1.1: Dataset description and precipitation
evaluation.
\newblock \emph{JAWRA Journal of the American Water Resources Association},
59(6):1246--1272, 2023.

\bibitem[Chung et~al.(2023)Chung, Kim, Mccann, Klasky, and Ye]{chung2023dps}
Hyungjin Chung, Jeongsol Kim, Michael~T. Mccann, Marc~L. Klasky, and Jong~Chul
Ye.
\newblock Diffusion posterior sampling for general noisy inverse problems.
\newblock In \emph{International Conference on Learning Representations}, 2023.

\bibitem[Hersbach et~al.(2020)Hersbach et~al.]{hersbach2020era5}
Hans Hersbach et~al.
\newblock The {ERA5} global reanalysis.
\newblock \emph{Quarterly Journal of the Royal Meteorological Society},
146(730):1999--2049, 2020.

\bibitem[Karras et~al.(2022)Karras, Aittala, Aila, and Laine]{karras2022edm}
Tero Karras, Miika Aittala, Timo Aila, and Samuli Laine.
\newblock Elucidating the design space of diffusion-based generative models.
\newblock In \emph{Advances in Neural Information Processing Systems}, 2022.

\bibitem[Kamath et~al.(2023)Kamath, Martilli, Singh, Brooks, Lanza, Bixler, Coudert, Yang, and Niyogi]{kamath2023h3i}
Harsh~G. Kamath, Alberto Martilli, Manmeet Singh, Trevor Brooks, Kevin Lanza,
R.~Patrick Bixler, Marc Coudert, Zong-Liang Yang, and Dev Niyogi.
\newblock Human heat health index (H3I) for holistic assessment of heat hazard
and mitigation strategies beyond urban heat islands.
\newblock \emph{Urban Climate}, 52:101675, 2023.
\newblock doi:10.1016/j.uclim.2023.101675.

\bibitem[Manshausen et~al.(2024)Manshausen, Cohen, Harrington, Pathak, Pritchard, Garg, Mardani, Kashinath, Byrne, and Brenowitz]{manshausen2024sda}
Peter Manshausen, Yair Cohen, Peter Harrington, Jaideep Pathak, Mike Pritchard,
Piyush Garg, Morteza Mardani, Karthik Kashinath, Simon Byrne, and Noah
Brenowitz.
\newblock Generative data assimilation of sparse weather station observations
at kilometer scales.
\newblock \emph{Journal of Advances in Modeling Earth Systems}, 17(10), 2025.

\bibitem[Rozet and Louppe(2023)]{rozet2023sda}
Fran\c{c}ois Rozet and Gilles Louppe.
\newblock Score-based data assimilation.
\newblock In \emph{Advances in Neural Information Processing Systems},
volume~36, pages 40521--40541, 2023.

\bibitem[Mardani et~al.(2025)Mardani, Brenowitz, Cohen, Pathak, Chen, Liu,
Vahdat, Nabian, Ge, Subramaniam, and Kashinath]{mardani2025corrdiff}
Morteza Mardani, Noah Brenowitz, Yair Cohen, Jaideep Pathak, Chieh-Yu Chen,
Cheng-Chin Liu, Arash Vahdat, Mohammad~Amin Nabian, Tao Ge, Akshay
Subramaniam, and Karthik Kashinath.
\newblock Residual corrective diffusion modeling for km-scale atmospheric
downscaling.
\newblock \emph{Communications Earth \& Environment}, 6(1):124, 2025.

\bibitem[Song et~al.(2021)Song, Sohl-Dickstein, Kingma, Kumar, Ermon, and
Poole]{song2021scoresde}
Yang Song, Jascha Sohl-Dickstein, Diederik~P. Kingma, Abhishek Kumar, Stefano
Ermon, and Ben Poole.
\newblock Score-based generative modeling through stochastic differential
equations.
\newblock In \emph{International Conference on Learning Representations},
2021.

\bibitem[Shepard(1968)]{shepard1968idw}
Donald Shepard.
\newblock A two-dimensional interpolation function for irregularly-spaced
data.
\newblock In \emph{Proceedings of the 1968 23rd ACM National Conference},
pages 517--524, 1968.

\bibitem[Luitel et~al.(2026)Luitel, Singh, Durkee, Fahad, Sudharsan, Singh,
He, Kamath, Yang, Halder, Juneja, Mukhopadhyay, Dhanuka, and
Srivastava]{luitel2026aircast}
Somnath Luitel, Manmeet Singh, Joshua Durkee, Abdullah~Al Fahad, Naveen
Sudharsan, Prabhjot Singh, Cenlin He, Harsh Kamath, Zong-Liang Yang,
Krishnagopal Halder, Sandeep Juneja, Parthasarathi Mukhopadhyay, Saptarishi
Dhanuka, and Amit~Kumar Srivastava.
\newblock {AirCast-SR}: A foundation model for kilometer-scale atmospheric
super-resolution via latent consistency diffusion.
\newblock \emph{arXiv preprint arXiv:2605.26130}, 2026.

\bibitem[Karras et~al.(2024)Karras, Aittala, Lehtinen, Hellsten, Aila, and
Laine]{karras2024edm2}
Tero Karras, Miika Aittala, Jaakko Lehtinen, Janne Hellsten, Timo Aila, and
Samuli Laine.
\newblock Analyzing and improving the training dynamics of diffusion models.
\newblock In \emph{Proceedings of the IEEE/CVF Conference on Computer Vision
and Pattern Recognition}, pages 24174--24184, 2024.

\bibitem[Kalnay(2002)]{kalnay2003atmospheric}
Eugenia Kalnay.
\newblock \emph{Atmospheric Modeling, Data Assimilation and Predictability}.
\newblock Cambridge University Press, 2002.

\end{thebibliography}
\end{document}